\documentclass[final,3p,times]{elsarticle}
\usepackage{amssymb}
\usepackage{amsmath}
\usepackage{multirow}
\usepackage{comment}
\usepackage{cite}
\usepackage{enumitem}
\usepackage{subcaption}
\usepackage{natbib}
\usepackage{xcolor}
\usepackage{booktabs}
\usepackage{array}
\usepackage{listings}
\usepackage{geometry}
\usepackage{float}
\usepackage{xcolor}

\journal{Journal of Systems and Software}

\begin{document}

\begin{frontmatter}

%% Title, authors and addresses

%% use the tnoteref command within \title for footnotes;
%% use the tnotetext command for theassociated footnote;
%% use the fnref command within \author or \affiliation for footnotes;
%% use the fntext command for theassociated footnote;
%% use the corref command within \author for corresponding author footnotes;
%% use the cortext command for theassociated footnote;
%% use the ead command for the email address,
%% and the form \ead[url] for the home page:
%% \title{Title\tnoteref{label1}}
%% \tnotetext[label1]{}
%% \author{Name\corref{cor1}\fnref{label2}}
%% \ead{email address}
%% \ead[url]{home page}
%% \fntext[label2]{}
%% \cortext[cor1]{}
%% \affiliation{organization={},
%%             addressline={},
%%             city={},
%%             postcode={},
%%             state={},
%%             country={}}
%% \fntext[label3]{}

\title{GateLens: A Reasoning-Enhanced LLM Agent for Automotive Software Release Analytics}

%% use optional labels to link authors explicitly to addresses:
%% \author[label1,label2]{}
%% \affiliation[label1]{organization={},
%%             addressline={},
%%             city={},
%%             postcode={},
%%             state={},
%%             country={}}
%%
%% \affiliation[label2]{organization={},
%%             addressline={},
%%             city={},
%%             postcode={},
%%             state={},
%%             country={}}

\author{Arsham Gholamzadeh Khoee\corref{cor1}\fnref{label1,label2}}
\ead{arsham.khoee@chalmers.se}
\author{Shuai Wang\fnref{label1,label2}}
\ead{shuaiwa@chalmers.se}
\author{Robert Feldt\fnref{label1}}
\ead{robert.feldt@chalmers.se}
\author{Dhasarathy Parthasarathy\fnref{label1}}
\ead{dhasarathy.parthasarathy@volvo.com}
\author{Yinan Yu\fnref{label1,label2}}
\ead{yinan@chalmers.se}
\cortext[cor1]{Corresponding Author}
%% Author affiliation
\affiliation[label1]{organization={Department of Computer Science and Engineering, Chalmers University of Technology},
            city={Gothenburg},
            country={Sweden}}

\affiliation[label2]{organization={Volvo Group},
            city={Gothenburg},
            country={Sweden}}

%% Abstract
\begin{abstract}
Ensuring reliable data-driven decisions is crucial in domains where analytical accuracy directly impacts safety, compliance, or operational outcomes. Decision support in such domains relies on large tabular datasets, where manual analysis is slow, costly, and error-prone. While Large Language Models (LLMs) offer promising automation potential, they face challenges in analytical reasoning, structured data handling, and ambiguity resolution. This paper introduces GateLens, an LLM-based architecture for reliable analysis of complex tabular data. Its key innovation is the use of Relational Algebra (RA) as a formal intermediate representation between natural-language reasoning and executable code, addressing the reasoning-to-code gap that can arise in direct generation approaches. In our automotive instantiation, GateLens translates natural language queries into RA expressions and generates optimized Python code. Unlike traditional multi-agent or planning-based systems that can be slow, opaque, and costly to maintain, GateLens emphasizes speed, transparency, and reliability. We validate the architecture in automotive software release analytics, where experimental results show that GateLens outperforms the existing Chain-of-Thought (CoT) + Self-Consistency (SC) based system on real-world datasets, particularly in handling complex and ambiguous queries. Ablation studies confirm the essential role of the RA layer. Industrial deployment demonstrates over 80\% reduction in analysis time while maintaining high accuracy across domain-specific tasks. GateLens operates effectively in zero-shot settings without requiring few-shot examples or agent orchestration. This work advances deployable LLM system design by identifying key architectural features—intermediate formal representations, execution efficiency, and low configuration overhead—crucial for domain-specific analytical applications where accuracy, traceability, and stakeholder trust are paramount.
\end{abstract}

%%Graphical abstract
%\begin{graphicalabstract}
%\includegraphics{grabs}
%\end{graphicalabstract}

%%Research highlights
%\begin{highlights}
%\item Research highlight 1
%\item Research highlight 2
%\end{highlights}

%% Keywords
\begin{keyword}
%% keywords here, in the form: keyword \sep keyword

%% PACS codes here, in the form: \PACS code \sep code

%% MSC codes here, in the form: \MSC code \sep code
%% or \MSC[2008] code \sep code (2000 is the default)
Large Language Models\sep Tabular Question Answering\sep Software Release Analytics\sep Automotive Software Testing\sep Test Result Interpretation\sep Interpretable Reasoning
\end{keyword}

\end{frontmatter}

%% Add \usepackage{lineno} before \begin{document} and uncomment 
%% following line to enable line numbers
%% \linenumbers

%% main text
%%

%% Use \section commands to start a section
\section{Introduction}
\begin{comment}
Validating software releases in the automotive industry is a multifaceted challenge, particularly for embedded software in safety-critical systems. Modern vehicles integrate numerous subsystems, making this process complex and resource-intensive. Each integration phase involves \emph{gating} steps---critical checkpoints where tests verify compliance with predefined quality standards. Failures at these gates can ripple through the system, delaying dependent subsystems regardless of their individual quality. Release managers tasked with safeguarding quality must analyze vast quantities of test results and validation data. While essential for ensuring safety and reliability, this process is time-consuming and prone to human error in data interpretation.
\end{comment}

% Yinan: Perhaps we can make it less automotive
Reliable decision-making in data-intensive domains depends on the ability to accurately analyze large volumes of structured data. In sectors such as automotive manufacturing, healthcare, finance, and regulatory compliance, critical decisions hinge on interpreting tabular datasets that capture test results, operational metrics, or validation records. These datasets often pass through gating steps—critical checkpoints where predefined quality or compliance standards must be met. Failures at these gates can cascade through interconnected processes, delaying dependent workflows regardless of their individual quality. Analysts tasked with safeguarding decision quality must process vast quantities of data. While essential for ensuring accuracy and reliability, this process is time-consuming and prone to human error in data interpretation.

The software industry's transition from manual to automated processes has entered a new era with the emergence of Large Language Models (LLMs)~\citep{10.3390/app14031046, chang2024survey}. Companies are increasingly integrating these AI agents into their workflows, seeking more cost-effective and optimized solutions for complex software engineering tasks~\citep{10.1109/ase56229.2023.00035}.
However, direct application of LLMs to structured data analysis for decision support is hindered by limitations in interpretable reasoning and understanding of technical specifications~\citep{10.3390/fi16060180, austin2021program}. 

\begin{figure*}[!h]
    \centering
    \includegraphics[width=1\linewidth]{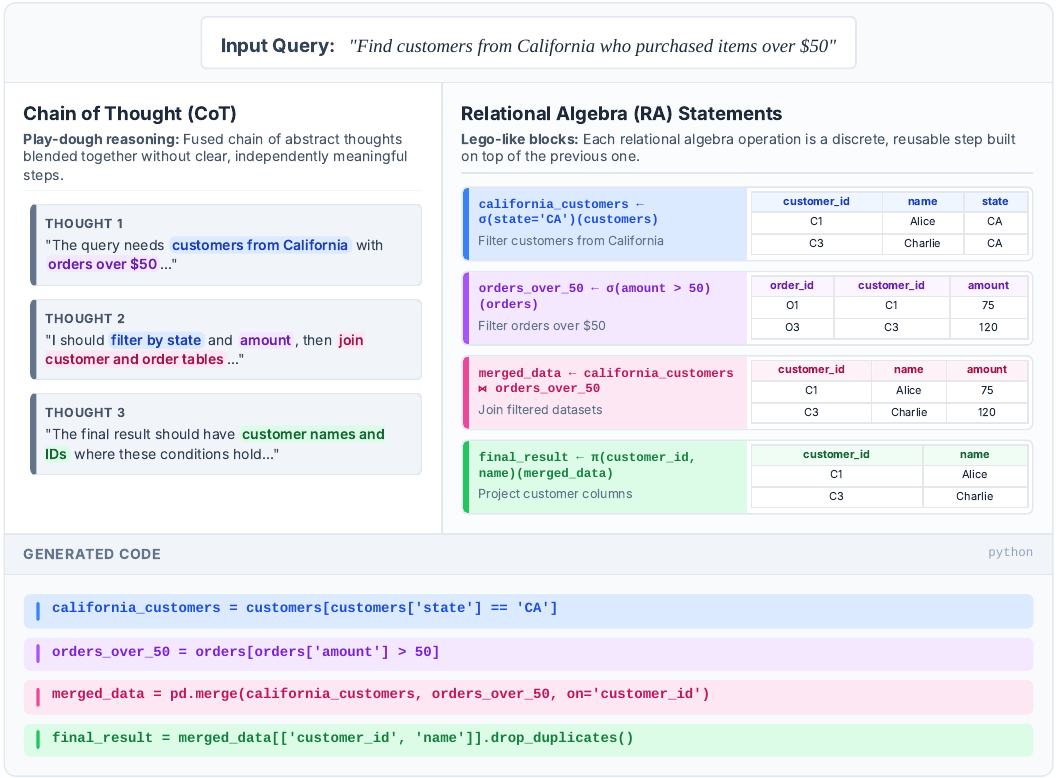} 

    \caption{Comparison of conventional Chain-of-Thought (CoT) reasoning and its extension using Relational Algebra (RA) statements. (Left) CoT employs unstructured, fused reasoning where multiple analytical concepts are blended together in informal thoughts without clear separation. Operations cannot be independently mapped to executable code snippets, making reasoning steps opaque and difficult to debug in isolation. (Right) RA-based reasoning adopts a structured, compositional approach where each relational algebra operation (filtering, joining, projecting) is a discrete, formally grounded step. Each operation is independently interpretable, reusable, and can be directly mapped to executable code. The color-coding illustrates how RA maintains clear boundaries between distinct analytical steps, whereas CoT conflates multiple operations within single reasoning thoughts. This structural distinction enables transparent, formally grounded reasoning that can be systematically verified and debugged at each stage.}
    \label{fig:cora}

\end{figure*}

To address these challenges, we introduce \emph{GateLens}, a reasoning-enhanced LLM agent~\citep{miehling2025agentic} for reliable tabular data analysis to support decision-making in domain-specific contexts. We validate the architecture in the automotive software release domain, where the need for precise, transparent, and efficient analysis is particularly acute.  

A key challenge in LLM-based analytical agents is the potential mismatch between natural-language reasoning and the actual computation implemented in generated code—a reasoning-to-code gap that becomes more pronounced as analytical complexity increases. GateLens addresses this challenge by integrating structured relational analysis with domain-specific expertise through a reasoning layer built on Relational Algebra (RA) as an extension of Chain-of-Thought (CoT). This reasoning layer systematically breaks down complex validation tasks into discrete, formally grounded analytical steps. We adopt this approach to address essential limitations of vanilla CoT for our work: its reasoning steps are opaque and often cannot be mapped directly to executable code, its operations are not compositional and cannot be independently reused or debugged, and its reasoning is unstructured with operations blended together without clear intent or formal grounding. In contrast, our RA-based reasoning layer aims to ensure that each step is independently interpretable and reusable, reasoning is grounded in formal relational algebra, and intent is made explicit through well-defined operations. Figure~\ref{fig:cora} illustrates this distinction between our structured RA-based approach and conventional CoT reasoning.

%GateLens integrates structured relational analysis with domain-specific expertise, leveraging a reasoning layer built on Relational Algebra (RA) to break down complex validation tasks into systematic analytical steps. %A domain-specific knowledge base---including automotive software specifications and data schemas---improves accuracy. 
%RF: Changed "ensures" to "improves"; the former sounds too strong??
GateLens simplifies three critical aspects of release validation: 

\noindent\textbf{1. Test Result Analysis:} Analyzing test execution outcomes is foundational to release validation. This involves analyzing pass/fail patterns across comprehensive test suites, identifying recurring failures, and validating test coverage metrics. In automotive software, where a single release might involve a large number of test cases across multiple vehicle functions, this task becomes particularly demanding. Release engineers must not only identify failed tests but also understand their patterns, assess coverage adequacy, and evaluate test execution stability.\\
\noindent\textbf{2. Impact Assessment:} Impact Assessment is a systematic process for evaluating how software issues affect vehicle functionality and safety during release validation. It involves three phases: first, a critical failure analysis identifies the root cause and immediate effects of an issue, such as an ABS module causing a 200ms brake signal delay that exceeds the 100ms threshold. Second, a component-level impact evaluation traces how the issue propagates through interconnected systems, assessing both direct effects, like problematic emergency braking, and indirect effects, such as reduced stability control performance. Finally, an integration risk assessment quantifies the severity of these impacts against safety thresholds and functional requirements, categorizing issues like the ABS delay as system-wide risks with critical severity. This structured process enables engineers to understand system-wide effects, ensuring all safety and functionality requirements are met before release.\\
% Shortened version: Engineers evaluate how software issues affect vehicle functionality and safety through systematic analysis of failure propagation across interconnected systems, assessing both direct and indirect effects against safety thresholds.
\noindent\textbf{3. Release Candidate Analysis:} The final quality gate involves evaluating Release Candidates (RCs) against predefined quality gates and criteria. This encompasses analyzing whether a particular RC meets all quality thresholds, identifying potential release blockers, and validating compliance with release requirements. In automotive software, where releases must meet stringent safety and quality standards, this analysis requires careful validation of each RC against established criteria, ensuring all prerequisites for a safe and reliable release are satisfied.

%Very short version of the entire part: \textbf{(1) Test Result Analysis} - identifying failure patterns and coverage gaps across test suites, \textbf{(2) Impact Assessment} - evaluating how software issues propagate through vehicle systems, and \textbf{(3) Release Candidate Analysis} - validating RCs against quality gates and safety requirements.

The traditional release validation process demands extensive manual effort. Release engineers meticulously analyze test results, assess impacts, verify RCs against quality gates, and report findings to stakeholders, such as release managers. As automotive software systems grow increasingly complex, these manual workflows become more challenging, time-consuming, and error-prone.

This work aims to streamline release validation by automating key analysis workflows, enabling engineers to focus on high-value analysis and discussion. By providing deeper analytical insights, the proposed approach reduces the time needed to deliver accurate validation results, empowering release managers to make informed decisions more efficiently.
Our \textbf{contributions} can be summarized as follows:
\begin{itemize}
\item We design an \emph{architecture optimized for time- and safety-critical environments}, minimizing LLM invocations while preserving reasoning depth for reliable tabular analysis. GateLens operates in a zero-shot setting, avoiding the need for few-shot examples or multi-agent coordination, which improves generalizability, execution speed, reduces maintenance overhead, and enhances transparency.

\item We introduce a \emph{scalable and maintainable framework for automotive software release validation}, developed in response to observed limitations in traditional planning-based multi-agent system. GateLens handles diverse user queries with higher robustness and clarity, supporting effective decision-making across a wider range of release engineering tasks.

\item We conduct a \emph{comprehensive empirical evaluation}, including comparisons with a CoT+SC system, ablation of the RA module, and performance across multiple LLMs (GPT-4o and Llama 3.1 70B). These experiments demonstrate GateLens's performance advantages in complex and ambiguous queries.

\item We report on \emph{real-world industrial deployment} in a partner automotive company, where GateLens reduces analysis time by over 80\% and demonstrates strong generalization across user roles, highlighting its practical value and deployment-readiness in safety-critical release validation workflows.
\end{itemize}

\section{Background and Motivation}

Software release decisions in the automotive industry involve multiple stakeholders and extensive data analysis. Modern vehicles integrate hundreds of software components, each requiring rigorous testing and validation. The process advances through distinct phases: component-level testing, integration testing, system-level validation, and vehicle validation testing. Component-level testing verifies individual software modules, integration testing ensures proper interaction between components, system-level validation examines the complete system behavior, and vehicle validation testing evaluates software performance under real vehicle conditions on closed tracks.

The development cycle grows in complexity with each integration phase. This increasing complexity presents challenges in managing large-scale test results, tracking interdependencies between components, correlating test failures across different subsystems, and maintaining historical context for recurring issues. The iterative nature of software testing and validation further expands this data ecosystem.

The wide range of stakeholders in the release process creates additional challenges in data interpretation and presentation. Project managers need high-level progress indicators, verification engineers require detailed technical insights, quality engineers focus on trend analysis and improvement metrics, and release engineers need specific release-readiness indicators. This variety of perspectives necessitates different views of the same underlying data, making the analysis process more complex.

This diversity is reflected not only in perspective but also in the level of granularity required from the underlying data. For example, senior management may ask high-level questions such as: ``List all vehicles that have not yet received global approval,'' seeking a consolidated overview without reference to schema details or subsystem-specific attributes. In contrast, release managers or verification engineers may pose highly specific queries such as: ``What are the baseline, phase, RC, and EUF values for RM-320 in the latest test suite?'' These questions rely on detailed knowledge of domain terminology and schema structure. Although both types of queries operate on the same data foundation, they demand substantially different views and levels of abstraction.

Release managers function as gatekeepers in the software deployment pipeline. They handle test result analysis, cross-system impact assessment, decision-making, stakeholder coordination, and safety compliance verification. The manual workflow introduces vulnerabilities: time-intensive processing, potential errors in interpretation, decision delays, and communication barriers between technical and business teams.

Within this process, statisticians provide an overall view of the data to project managers and quality engineers for future business decisions. The existing manual approach faces several limitations, particularly regarding time and resource constraints. These include labor-intensive data analysis, delayed response to critical issues, limited capacity for comprehensive analysis, and bottlenecks in the release pipeline. Communication challenges further complicate the process, with misalignment between technical analysts and statisticians, varying interpretations of project requirements, inconsistent reporting formats, and knowledge transfer gaps.

Internal Testing on Closed Track represents a crucial validation phase. Release managers must analyze extensive datasets to evaluate progression readiness. The manual query process for report generation can impact release timelines, business objectives, and subsystem integration schedules.

The deployment of intelligent assistants presents opportunities to address these challenges through automated data processing and analysis, standardized reporting frameworks, real-time insights generation, and stakeholder-specific view generation. However, current automation solutions, including basic LLM implementations, face limitations in understanding complex technical specifications, maintaining structured analysis steps, handling domain-specific requirements, and processing automotive validation data systematically.

These limitations highlight the need for enhanced solutions combining domain expertise with advanced analytical reasoning capabilities. Such solutions must facilitate efficient decision-making while maintaining high safety and quality standards in automotive software development~\citep{wang2024surveyl}. Effective intelligent assistants can transform the release decision process, enabling release managers to prioritize result interpretation and strategic decision-making over routine data analysis.

\section{Approach and Methodology}
The complexity of automotive software release validation demands a system that can bridge the gap between human-centric inquiries and precise technical analysis. GateLens addresses this challenge by utilizing LLM agents to transform natural language queries into actionable insights through systematic analysis.
At its core, GateLens must fulfill three fundamental requirements:

\noindent\textbf{1. Query Understanding:} The system must accurately interpret diverse user queries, ranging from high-level management questions to detailed technical inquiries about specific test cases.\\
\noindent\textbf{2. Query Transformation:} The system needs to transform these interpretations into structured formal expressions, ensuring consistent and verifiable reasoning structures.\\
\noindent\textbf{3. Analysis Execution:} The system must generate and execute precise analysis code that processes validation data according to these formal expressions.

The architecture of GateLens is driven by these fundamental requirements, establishing a systematic pipeline for transforming user queries into analytical results. The system leverages RA to enhance LLMs' reasoning capabilities and transform user queries into formal relational operations. To support this transformation, GateLens employs domain-specific data schemas that guide its relational modeling and enhance the generation of RA expressions. This approach ensures both accessibility for non-technical stakeholders and the precision required for automotive software validation.

\begin{figure*}[ht]
    \centering
    \includegraphics[width=1.02\linewidth]{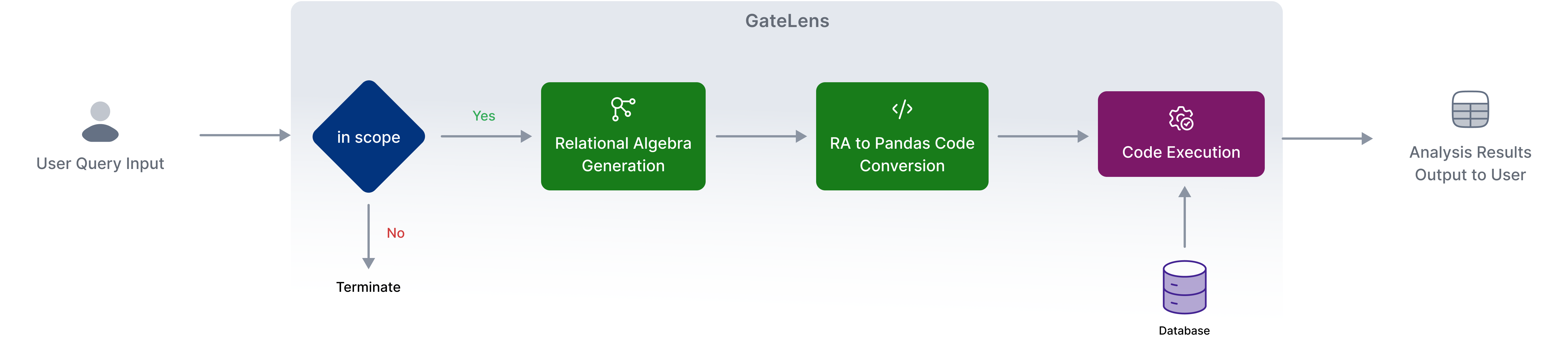} 

    \caption{GateLens top-level architecture: The system processes high-level queries from the end user, generates the necessary data manipulation code using the enhanced reasoning layer with the help of RA, executes it, and outputs the result table as a decision-support resource.}
    \label{fig:architecture}

\end{figure*}

\subsection{System Overview}
The primary objective of GateLens is to generate executable code that performs precise test data analysis based on user queries. The system's workflow consists of two main phases: query interpretation and code generation. As illustrated in Figure~\ref{fig:architecture}, GateLens first processes user queries through an LLM agent that translates natural language inputs into formal RA expressions. This translation incorporates a detailed relational model of the test data, ensuring precise specification of automotive domain concepts. 
The resulting RA expressions serve as a pivotal intermediate representation that is more transparent to both LLM agents and humans. 
In the second phase, these formal expressions are passed to the coder agent, which generates executable desirable code, such as SQL or Python code, to perform the required analysis on the test data and produce results.

\subsection{Core Components}
\label{subsec:core}
\begin{figure}[ht]
    \centering
    \includegraphics[width=0.6\linewidth] {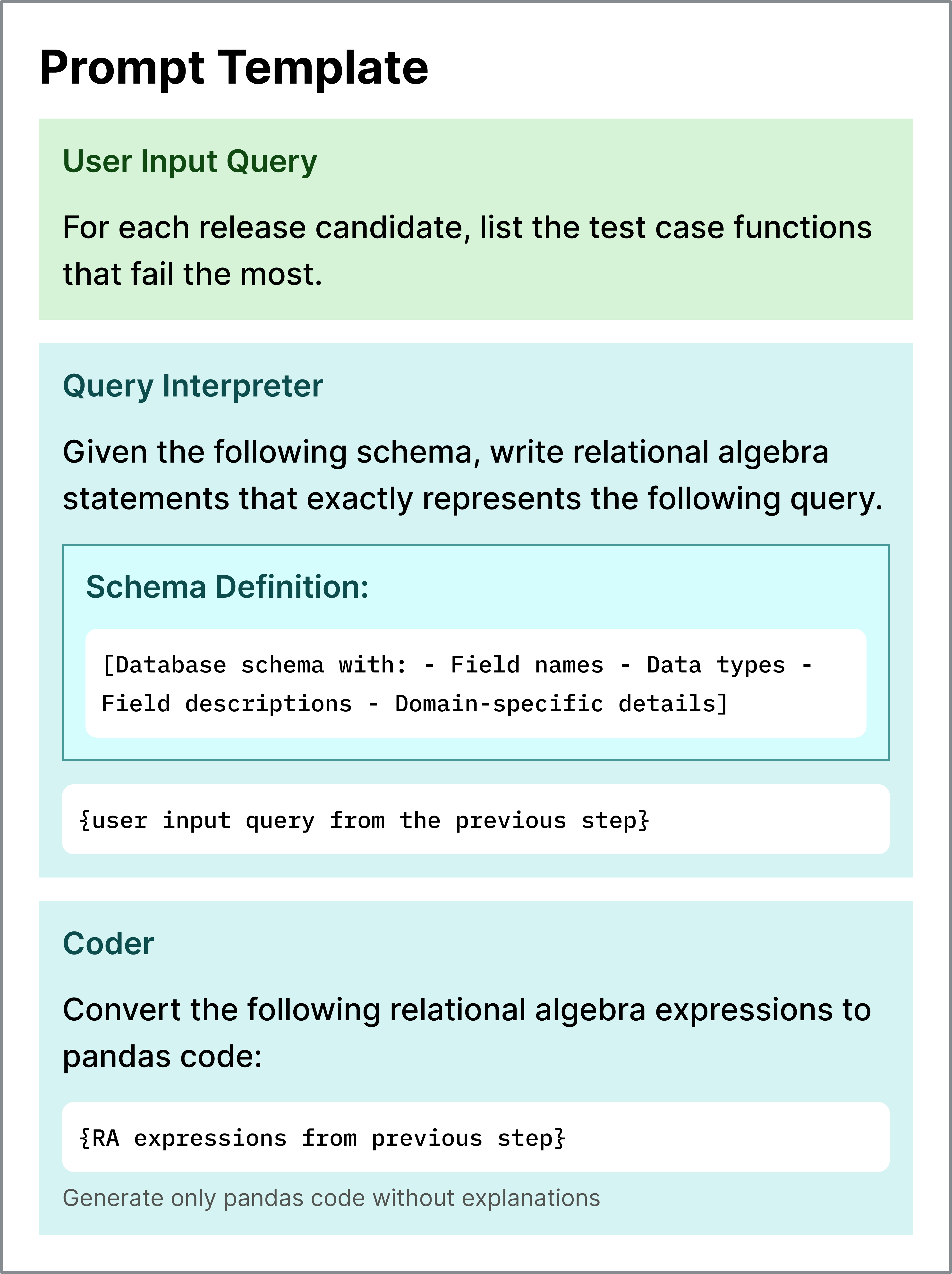} 
    \caption{Overview of the prompt engineering flow and prompt structure within the GateLens system architecture.}
    \label{fig:prompts}

\end{figure}

The system architecture consists of two primary components that work in tandem to transform natural language queries into executable code: the query interpreter agent and the coder agent. The query interpreter agent first translates user queries into RA expressions, providing a structured framework for analytical reasoning. The coder agent then converts these formal expressions into executable code, completing the transformation pipeline. This two-stage approach ensures both analytical precision and efficient implementation, where the prompt engineering flow and prompt structure are presented in Figure \ref{fig:prompts}.

\subsubsection{Query Interpreter}
The query interpreter agent is responsible for converting user queries into formal RA expressions, providing a precise framework for analytical reasoning. Before initiating this translation, the agent consults the knowledge base, comprising the data schema and domain-specific context. The data schema provides a detailed understanding of the dataset, including its relational modeling, field descriptions, data types, and enriched metadata capturing domain-specific acronyms and terminology mappings. This glossary-enhanced schema is injected into the prompt context, enabling accurate resolution of domain-specific terms into formal schema attributes during RA construction. 

To handle imprecise queries without compromising data privacy or exceeding LLM context windows, GateLens employs a selective strategy for exposing categorical information. For low-cardinality attributes (e.g., fields with only a few distinct categories such as test status), all valid options are explicitly enumerated within the schema metadata. In contrast, for high-cardinality attributes, actual database values are never exposed; instead, the schema specifies structural patterns or expected formats (e.g., standard prefixes or identifier conventions). This hybrid design keeps the prompt context compact and privacy-preserving while providing the LLM with sufficient context to generate accurate filtering conditions.

Using this information, the agent verifies whether the query is relevant and within the scope of the dataset~\citep{manik2021out}. This validation step ensures that only supported and meaningful queries are processed, improving both accuracy and efficiency. Once the query is confirmed to be in scope, the agent leverages the data schemas to interpret and decompose the query into formal RA expressions.

The agent's primary function is to map natural language queries into formal RA expressions, enhancing LLM reasoning through structured decomposition~\citep{khot2022decomposed}. This approach extends traditional Chain-of-Thoughts (CoT)~\citep{wei2022chain} reasoning by constraining the model to think within a formal system framework~\citep{zhang2023igniting}. Instead of generating free-form solutions, the agent must express analytics using standard relational algebra notations through a limited set of standard operations: selection, projection, union, set difference, cartesian product, and rename as basic operations, as well as derived operations such as join, intersection, and division and complemented by aggregation functions like average, minimum, maximum, sum, and count.

By limiting operations to this standard set, the agent effectively handles ambiguous queries through formal translation, ensures technical precision, and prevents deviation from analytical requirements. The formal nature of RA enables query optimization, which the agent incorporates by prioritizing data reduction operations early in the expression chain. This optimization strategy involves applying filters first, then performing expensive operations on the reduced dataset, thereby minimizing processing time and resource utilization.

The translation to RA offers two significant advantages. First, it makes the analytics more transparent in technical terms, allowing for clear interpretation and validation of the reasoning process. Second, it ensures that every solution generated is precisely defined and feasible for implementation, preventing the agent from proposing impractical or undefined analytical approaches.

\subsubsection{Coder}

The coder agent is responsible for generating executable code from given RA expressions. Upon receiving an RA expression, the agent follows precise instructions to produce code that delivers the final analytical results. This capability allows the agent to generate complete, self-contained code at once, eliminating the need for step-by-step generation and execution phases.

To ensure the generated code meets quality standards, we designed prompts that specifically instruct the LLM to include essential validation mechanisms. The prompts explicitly require data type validation instructions for numerical and categorical field handling, null value checking procedures to maintain data integrity, validation of numerical operations, and validation of join conditions to ensure proper matching of key columns between tables. 

By generating the entire code in a single pass rather than through iterative refinement, the agent significantly reduces processing overhead, system response time, and resource consumption while minimizing potential errors that could arise from multiple execution steps. This streamlined yet precise approach ensures both efficiency and reliability in the analysis pipeline, fully aligned with the formal rigor of RA expressions and improving overall system responsiveness to user queries.

\subsection{Data Handling}
A key architectural decision in GateLens is its indirect interaction with test data. Rather than exposing sensitive test data directly to LLM agents, which could raise privacy concerns~\citep{boudewijn2023privacy} and exceed input context limitations, the system operates on data schemas and relational models. This approach serves multiple critical purposes:
\begin{itemize}
    \item \textbf{Privacy Protection:} Sensitive automotive test data remains secure within the organization's infrastructure
    %\item \textbf{Hallucination Prevention:} By limiting LLMs to generating analytical code from data schemas rather than processing raw data directly, the system prevents potential hallucinations and incorrect inferences about actual data values
    \item \textbf{Scalability:} The system can handle large-scale test datasets that would exceed LLM context windows
    \item \textbf{Knowledge Integration:} Data schemas and relational models serve as an essential knowledge base, providing necessary structural understanding without raw data exposure
\end{itemize}
The final execution of the generated code runs on the test data in the target environment, maintaining data privacy while delivering precise analytical results.

\section{Experimental Evaluation}
\label{sec:experiments}
In this section, we aim to answer the following research questions:

\begin{enumerate}[label=RQ\arabic*:]
\item  How effectively does GateLens address user queries and deliver accurate results across various query categories?

\item  How robust is GateLens in handling out of scope queries and imprecise queries?

\item  How does the RA reasoning procedure contribute to the overall performance of GateLens?

\item  To what extent does RA-based reasoning eliminate the need for in-context learning?

% \item  What benefits does GateLens provide to different roles in real-world industrial contexts? 
\end{enumerate}

\subsection{Experimental Setup}

To address the research questions introduced in Section~\ref{sec:experiments}, we designed and conducted extensive experiments to evaluate the performance of GateLens.

The experimental data comprises two distinct benchmarks. The first benchmark consists of 50 queries designed with the assistance of release engineers, quality engineers, and verification engineers. To assess GateLens’s performance across a spectrum of query complexities, these queries are categorized into four difficulty levels. The four levels of query difficulty are defined as follows:

\noindent\textbf{\underline{Level 1}} Simple queries involving a single operation such as filtering or sorting.\\
\noindent\textbf{\underline{Level 2}} Queries combining two or three basic operations, such as multiple filtering followed by sorting.\\
\noindent\textbf{\underline{Level 3}} Queries involving more than three operations, potentially including grouping and aggregating.\\ 
\noindent\textbf{\underline{Level 4}} Complex queries requiring multiple advanced operations beyond basic filtering and sorting, such as grouping and aggregating for statistical calculations.

The second benchmark is derived from real-world user queries collected from production logs at our partner company. These queries were sourced from the historical logs of an agentic system that employed a well-established tabular data reasoning approach combining  CoT~\citep{wei2022chain} prompting with Self-Consistency (SC)~\citep{wang2022self}. This system was used in production to support software release analytics, and the collected queries reflect a wide range of user roles, query types, and domain-specific requirements.
While this system performs effectively in many scenarios, its limitations become apparent as the range of roles and users expands, leading to a significant diversification of queries. This broader query diversity exposes the system’s reliance on few-shot examples, making it less capable of handling highly complex, ambiguous, or ill-defined queries that require greater flexibility and adaptability. Nevertheless, this preliminary system played a critical role in data collection for GateLens by providing query logs used to develop and validate our approach. From these logs, we filtered out near-duplicates and selected 244 frequently repeated unique queries, which we then organized into eight functional categories based on their purposes. The ground truth for the 244 real-world queries was established through a two-stage manual annotation process: two domain experts independently solved an initial subset of 20 queries to align on interpretation and expected outputs, after which the remaining queries were divided and annotated following the agreed-upon criteria, with periodic cross-checks to ensure consistency.

\iffalse
The second benchmark is derived from real-world user queries collected from production logs. These queries were curated from the historical logs of the first-generation system, GoNoGo, a multi-agent system designed to support software release decision-making using planning and few-shot learning, which is currently deployed in production at our partnering industrial company~\citep{khoee2024gonogo}. 
While GoNoGo performs effectively in many scenarios, its limitations become evident as the range of roles and users expands, leading to a significant diversification of queries. This increased diversity exposes GoNoGo's reliance on few-shot examples, making it less capable of handling highly complex, ambiguous, or ill-defined queries that require greater flexibility and adaptability. Despite its limitations, the GoNoGo system played a critical role in data collection for GateLens by providing query logs that aided in developing and validating future iterations. Leveraging these historical logs, we filtered out very similar queries and selected 244 of the most frequently repeated unique queries. These queries were then categorized into eight functional categories based on their purposes.
\fi

In order to assess GateLens’s performance, we run experiments with two large language models (LLMs): GPT-4o, a leading commercial model, and Llama 3.1 70b, a recently released open-source model. We also benchmark GateLens against the CoT+SC~\citep{khoee2024gonogo} agentic system currently used to support the company’s release decisions. Our comparative analysis is designed to quantify the improvements introduced by GateLens’s novel architecture.

\iffalse
In order to assess GateLens's performance, a series of experiments are conducted using two large language models (LLMs): GPT-4o, a leading commercial model in its class, and Llama 3.1 70b, a recently introduced open-source model. GateLens is also compared with the latest version of the first-generation system, GoNoGo \citep{khoee2024gonogo}, our recently deployed production system that has demonstrated significant value in industrial settings. While GoNoGo effectively serves the company's release engineering needs, this comparative analysis aims to evaluate potential improvements in our new system design. 
It is important to note that this version of GoNoGo differs slightly from the one introduced in the original paper, as it incorporates minor updates in prompt engineering and replaces GPT-3.5 with the aforementioned recent advanced LLMs, while still maintaining its core approach based on Chain-of-Thought (CoT)~\citep{wei2022chain} reasoning and Self-Consistency~\citep{wang2022self} techniques.
\fi

To address the challenge of handling out-of-scope user queries during real-time interactions, GateLens incorporates an in-scope filtering mechanism as explained in Section~\ref{subsec:core}. This mechanism ensures that the system only attempts to process queries that fall within its scope, thereby improving reliability and reducing errors.
Performance evaluation focused on two key aspects:
\begin{enumerate}
    \item \textbf{Quality of responses}: Measured using precision, recall, and F1 Score, which reflect the system's ability to address relevant queries correctly.
    \item \textbf{Coverage of relevant queries}: Ensuring the system does not reject a significant proportion of valid queries, thus maintaining broad applicability.
\end{enumerate}

Additionally, an ablation study is conducted to examine the contribution of the RA reasoning mechanism of GateLens. 

In our experiments, the evaluation of system performance is based on the following definitions: A \textbf{True Positive (TP)} occurs when the system produces a result that matches the manually generated ground-truth result, specifically when the final output of the executed code matches the expected ground-truth output. A \textbf{False Positive (FP)} occurs when the system provides an incorrect result, meaning the executed code produces output that differs from the ground-truth. A \textbf{False Negative (FN)} occurs when the system fails to provide any result for a query. \textbf{True Negatives (TNs)} are not applicable since we focus on valid queries producing meaningful output.

Based on these definitions, we calculate Precision, Recall, and F1 scores to assess the performance of the system. Precision ensures that incorrect results are minimized, recall ensures relevant queries are addressed, and the F1 score balances the two to provide an overall assessment of system performance. By relying on a closed-set benchmark with established ground truths, these metrics enable us to rigorously isolate and measure the impact of our architectural choices during lab validation, setting the stage for the real-world industrial evaluation detailed in Section~\ref{sec:lessons}.

\subsection{Performance in Addressing User Queries (RQ1)}
\label{subsec:rq1}

\begin{table*}[h!]
\centering
\caption{Comparison of average token consumption and reduction between CoT+SC and GateLens across four difficulty levels. The evaluation was conducted on the first benchmark, which consists of 50 designed queries with annotated difficulty levels, with both agents utilizing GPT-4o.}

\resizebox{1\textwidth}{!}{
\begin{tabular}{|c|c|ccc|ccc|ccc|ccc|}
\hline
\multirow{2}{*}{\textbf{Level}} & \multirow{2}{*}{\textbf{\# Queries}} & \multicolumn{3}{c|}{\textbf{GateLens with GPT-4o}} & \multicolumn{3}{c|}{\textbf{GateLens with Llama 3.1 70B}} & \multicolumn{3}{c|}{\textbf{CoT+SC with GPT-4o}} & \multicolumn{3}{c|}{\textbf{CoT+SC with Llama 3.1 70B}} \\ \cline{3-14} 
 &  & \textbf{Precision} & \textbf{Recall} & \textbf{F1 Score} & \textbf{Precision} & \textbf{Recall} & \textbf{F1 Score} & \textbf{Precision} & \textbf{Recall} & \textbf{F1 Score} & \textbf{Precision} & \textbf{Recall} & \textbf{F1 Score} \\ \hline
1 & 16 & 100\% & 100\% & \textbf{100\%} & 100\% & 43.75\% & 60.87\% & 93.33\% & 87.5\% & 90.32\% & 100\%      & 93.75\%    & 96.77\%\\ \hline
2 & 16 & 100\% & 100\% & \textbf{100\%} & 100\% & 62.5\% & 76.92\% & 100\% & 81.25\% & 89.66\% & 92.31\%    & 75\%       & 82.76\%\\ \hline
3 & 12 & 100\% & 100\% & \textbf{100\%} & 100\% & 50\% & 66.67\% & 91.67\% & 91.67\% & 91.67\% & 90.91\%    & 83.33\%    & 86.96\%\\ \hline
4 & 6 & 100\% & 100\% & \textbf{100\%} & 100\% & 33\% & 49.62\% & 66.67\% & 66.67\% & 66.67\% & 60\%       & 50\%       & 54.55\%\\ \hline
\textbf{Total} & \textbf{50} & \textbf{100\%} & \textbf{100\%} & \textbf{100\%} & \textbf{100\%} & \textbf{47.31\%} & \textbf{63.52\%} & \textbf{87.91\%} & \textbf{81.77\%} & \textbf{84.57\%} & \textbf{85.81\%}       & \textbf{75.52\%} & \textbf{80.26\%}\\ \hline
\end{tabular}
}
\label{tab:performance_comparison_gt50}
\end{table*}

\begin{table*}[h!]
\centering
\caption{Performance comparison of GateLens and the CoT+SC system across different categories on the second benchmark, which consists of 244 real-world queries.}

\resizebox{1\textwidth}{!}{
\begin{tabular}{|l|c|ccc|ccc|ccc|}
\hline
\multirow{2}{*}{\textbf{Category}} & \multirow{2}{*}{\textbf{\# Queries}} & \multicolumn{3}{c|}{\textbf{GateLens with GPT-4o}} & \multicolumn{3}{c|}{\textbf{GateLens with Llama 3.1 70B}} & \multicolumn{3}{c|}{\textbf{CoT+SC with GPT-4o}} \\
\cline{3-11}
 &  & \textbf{Precision} & \textbf{Recall} & \textbf{F1 Score} & \textbf{Precision} & \textbf{Recall} & \textbf{F1 Score} & \textbf{Precision} & \textbf{Recall} & \textbf{F1 Score} \\
\hline
Column Operations & 17 & 64.7\% & 64.7\% & 64.7\% & 50\% & 11.76\% & 19.04\% & 76.47\% & 76.47\% & \textbf{76.47}\% \\
Complex Multi-Condition Queries & 77 & 86.3\% & 81.82\% & \textbf{84\%} & 100\% & 24.68\% & 39.59\% & 84.75\% & 64.94\% & 73.53\% \\
Conditional Calculations & 8 & 100\% & 87.5\% & \textbf{93.3\%} & 100\% & 37.5\% & 54.55\% & 87.5\% & 87.5\% & 87.5\% \\
Data Filtering & 32 & 89.66\% & 81.25\% & \textbf{85.25\%} & 90.91\% & 31.25\% & 46.51\% & 86.21\% & 78.13\% & 81.97\% \\
Duplicate Removal & 78 & 87.67\% & 82.05\% & \textbf{84.77\%} & 100\% & 23.08\% & 37.5\% & 75.93\% & 52.56\% & 62.12\% \\
Grouping and Aggregation & 10 & 80.0\% & 80.0\% & \textbf{80.0\%} & 100\% & 30\% & 46.15\% & 83.33\% & 50\% & 62.5\% \\
Metadata Queries & 13 & 91.67\% & 84.61\% & \textbf{88.0\%} & 100\% & 15.38\% & 26.66\% & 46.15\% & 46.15\% & 46.15\% \\
Table Generation & 9 & 88.89\% & 88.89\% & \textbf{88.89\%} & 100\% & 44.44\% & 61.53\% & 88.89\% & 88.89\% & \textbf{88.89\%} \\
\hline
\textbf{Total} & \textbf{244} & \textbf{86.02\%}       & \textbf{81.14\%}       & \textbf{83.51\%}  & \textbf{92.61\%} & \textbf{27.26\%} & \textbf{41.44\%} & \textbf{83.15\%} & \textbf{63.52\%} & \textbf{70.61\%} \\
\hline
\end{tabular}
}

\label{tab:performance_comparison_user244}
\end{table*}

We conducted experiments to compare the performance of GateLens across the two introduced benchmarks. The first benchmark, consisting of 50 queries categorized by difficulty levels, was used to evaluate and compare the performance of GateLens and CoT+SC. Both systems were tested using GPT-4o and Llama 3.1 70B as their underlying LLMs. The results are summarized in Table~\ref{tab:performance_comparison_gt50}.

The results demonstrate that GateLens with GPT-4o significantly outperforms GateLens with Llama 3.1 70B, indicating GPT-4o’s superior capability for interpreting and generating RA. Similarly, CoT+SC with GPT-4o outperforms its Llama 3.1 70B variant, with the performance gap growing as query complexity increases. CoT+SC performance declines with query complexity. This underscores the importance of the RA reasoning mechanism in GateLens, which enables effective handling of complex, unstructured queries by decomposing them into logical, structured expressions. Most notably, GateLens with GPT-4o achieved optimal performance on this benchmark, maintaining 100\% accuracy across all difficulty levels. This stems from integrating RA reasoning into our framework. By translating queries into RA expressions, GateLens explicitly captures the logical structure of operations, enhancing both the clarity and precision of the generated code. The intermediate RA conversion allows the system to focus on the relevant table operations while filtering out irrelevant elements in the query, greatly enhancing the problem-solving capabilities of the LLM agent.

For the second benchmark, results in Table~\ref{tab:performance_comparison_user244} show that GateLens (GPT-4o) and CoT+SC (GPT-4o) significantly outperformed GateLens with Llama 3.1 70B. This performance disparity is primarily due to the strict code generation requirements of the task, including table filtering, merging strategies, and key-value mapping operations, where GPT-4o demonstrated markedly superior capabilities.

GateLens with GPT-4o outperformed CoT+SC (GPT-4o) across most categories, particularly evident in Metadata Queries (those seeking basic table information). For example, when processing the query "Give me the list of release candidates," the CoT+SC system often fails to identify the correct field. A common failure mode in CoT+SC occurred when user queries included typographical errors or incorrect casing in field names, with the system directly using the erroneous fields without correction. GateLens addresses this limitation through its query-to-RA transformation process, which incorporates the database's relational model, adjusts query fields to match table formats, and can handle fuzzy matching to detect and correct field names, enabling the system to resolve typographical errors and ambiguous queries effectively. This approach improves accuracy and resilience, particularly in real-world scenarios where user queries may not always adhere to strict formatting standards.

\begin{table*}[h!]
\centering
\caption{Comparison of average token consumption and reduction between CoT+SC and GateLens across four difficulty levels. The evaluation was conducted on the first benchmark, which consists of 50 designed queries with annotated difficulty levels, with both agents utilizing GPT-4o.}
\label{tab:token_consumption}
\resizebox{1\textwidth}{!}{
\begin{tabular}{|c|l|r|r|r|c|}
\hline
\textbf{Level} & \textbf{Agent} & \textbf{Avg Input Tokens} & \textbf{Avg Output Tokens} & \textbf{Avg Total Tokens} & \textbf{Token Reduction} \\ \hline
\multirow{2}{*}{1} 
 & CoT+SC & 11,905 & 186 & 12,091 & --- \\ 
 & GateLens & 2,239 & 420 & 2,658 & $\downarrow$ 78\% \\ \hline
\multirow{2}{*}{2} 
 & CoT+SC & 13,747 & 368 & 14,116 & --- \\ 
 & GateLens & 2,428 & 701 & 3,129 & $\downarrow$ 78\% \\ \hline
\multirow{2}{*}{3} 
 & CoT+SC & 14,726 & 441 & 15,168 & --- \\ 
 & GateLens & 2,432 & 698 & 3,130 & $\downarrow$ 79\% \\ \hline
\multirow{2}{*}{4} 
 & CoT+SC & 17,103 & 452 & 17,555 & --- \\ 
 & GateLens & 2,505 & 847 & 3,352 & $\downarrow$ 81\% \\ \hline
\end{tabular}
}
\label{tab:token_comparison}
\end{table*}

In addition, compared to the CoT+SC solution, GateLens uses significantly fewer tokens due to its effective use of RA as the intermediate representation and its zero-shot architecture. In Table~\ref{tab:token_comparison}, we compare the token usage of both systems across different difficulty levels. Notably, while the CoT+SC approach requires increasingly massive input contexts, primarily to accommodate few-shot examples, GateLens maintains a highly compact input footprint. As the query difficulty increases, the token reduction becomes even more significant, demonstrating the efficiency and scalability of the RA-based approach in handling complex queries.

\vspace{0.25cm}
\noindent\fbox{%
    \parbox{0.98\linewidth}{%
        \textbf{RQ1 findings:}
        GateLens with GPT-4o achieved 100\% F1 score on the first benchmark across all difficulty levels and 83.51\% F1 score on the second benchmark with 244 real-world queries. It outperformed CoT+SC (GPT-4o) by approximately 13 percentage points on real-world queries, indicating that the RA reasoning mechanism effectively addresses both complex logical operations and real-world noise, such as typos and ambiguous field names.
    }%
}

\subsection{Robustness: Handling Out of Scope and Imprecise Queries (RQ2)}

To assess the robustness of our approach in handling diverse user queries under real-world conditions, we conducted further experiments focusing on filtering out-of-scope queries as well as processing imprecise queries. For this purpose, the data analysis team at our industrial partner company manually selected 37 out-of-scope queries and 50 imprecise queries from the historical logs of the first-generation system, which are used to perform targeted evaluations.

Out-of-scope queries are those that cannot be meaningfully answered using the available data. For example, a query like "What is the most beautiful truck?" requires subjective judgment and cannot be resolved through database operations; it should be identified and filtered as out of scope. On the other hand, imprecise queries are those that can be answered using the database but contain ambiguous or inexact terms. For instance, a query such as "Find some trucks for cases that are NOK" is considered imprecise because while it seeks truck names where test results are "NOK" (failed), it uses ambiguous terminology - referring to "trucks" instead of the actual database field "name", and mentions "NOK" without specifying the "test\_result" field. Such imprecise queries require mapping informal language to precise database fields and conditions for proper execution.

% For instance, a query such as "List 10 trucks " corresponded to a column labeled "name" in the table. However, the term "truck" did not explicitly match any header, introducing ambiguity for query interpretation and code generation.

\subsubsection{Handling Out of Scope Queries}

We compared GateLens with other models; the results can be found in Table~\ref{tab:unrelated_query}. The results demonstrate that GateLens with GPT-4o achieved the best performance, particularly in terms of precision, which is approximately 40\% higher than other models, indicating GateLens' ability to avoid generating incorrect results.

The superior precision of GateLens with GPT-4o can be attributed to two key aspects of its design. First, its robust filtering mechanism ensures that out-of-scope queries are identified and excluded early in the processing pipeline, preventing irrelevant results. Second, the conversion of raw natural language queries into structured RA expressions enables the model to isolate and capture task-relevant components of a query. This structured approach considerably decreases erroneous outcomes and enhances the model's ability to handle complex and diverse query formulations in real-world scenarios.

\begin{table}[ht]
\centering
\caption{Model comparison for out-of-scope queries.}

\resizebox{0.48\textwidth}{!}{
\begin{tabular}{|c|c|c|c|}
\hline
\textbf{Model}       & \textbf{Precision}    & \textbf{Recall}     & \textbf{F1 Score}    \\ \hline
\textbf{GateLens with GPT-4o}       & 92.5\%         & 100\%          & \textbf{96.10\%}       \\ \hline
\textbf{GateLens with Llama 3.1 70B} & 52.94\%        & 97.30\%        & 68.57\%       \\ \hline
\textbf{CoT+SC with GPT-4o}      & 51.10\%        & 89.19\%        & 64.97\%       \\ \hline
\end{tabular}
}

\label{tab:unrelated_query}
\end{table}

CoT+SC showed significantly lower precision due to the variability of real-world queries and the inconsistency of user narratives, which often contain a mix of relevant and irrelevant content. This variability increases uncertainty and poses challenges for models that struggle to identify task-relevant information. Although all models demonstrated high recall, this did not translate into accurate processing. 
\begin{comment}
In contrast, the precision of CoT+SC was notably weaker. This can be attributed to the diversity of real-world queries and the variability in user narratives, which often contain a mixture of relevant and irrelevant parts. Such queries increase the degree of uncertainty, making it challenging for models that cannot effectively isolate task-relevant parts to process them appropriately. Furthermore, while the recall rates of all models were relatively high, indicating their ability to generate some results, this does not necessarily translate to accurate query processing.

CoT+SC showed a significant limitation, which relies on few-shot learning, as it occasionally generated code for out-of-scope queries. This behavior likely stems from the task-relevant examples in its prompt, which inadvertently bias the model toward attempting code generation even for out-of-scope queries. In summary, our approach demonstrates a strong ability to filter out-of-scope queries while maintaining a low error rate when processing in-scope queries.
\end{comment}

\subsubsection{Handling Imprecise Queries}

To further assess the robustness of our approach, we evaluated its performance in handling imprecise queries, which posed two primary challenges. First, some queries are informal and conversational in style, appearing unrelated to data analysis but actually carrying relevant intent. Second, many queries referred to fields using terms differing from the column headers.

\begin{table}[t]
\centering
\caption{Model comparison for imprecise queries.}

\resizebox{0.48\textwidth}{!}{
\begin{tabular}{|c|c|c|c|}
\hline
\textbf{Model} & \textbf{Precision}    & \textbf{Recall}    & \textbf{F1 Score}    \\ \hline
\textbf{GateLens with GPT-4o}     & 92.86\%       & 78\%          & \textbf{84.78\% }       \\ \hline
\textbf{GateLens with Llama 3.1 70B} & 92.86\%      & 26\%          & 40.63\%        \\ \hline
\textbf{CoT+SC with GPT-4o}    & 90\%          & 36\%          & 51.43\%        \\ \hline
\end{tabular}
}

\label{tab:fuzzy_query}
\end{table}

The results of these experiments are presented in Table \ref{tab:fuzzy_query}. As shown, GateLens with GPT-4o demonstrates the best overall performance. In terms of precision, all methods performed relatively well, indicating that when results are generated, they are likely to be correct. However, our method significantly outperformed the others in recall, highlighting its ability to handle a larger portion of the imprecise queries. As a result, GateLens with GPT-4 achieved a substantially higher F1 score compared to other methods, demonstrating that it not only processes most queries but also produces accurate results for them.

The observed performance gap between GateLens with GPT-4o and the other models can be attributed to their inherent limitations. Specifically, the Llama 3.1 70B model struggled to interpret user queries that deviated from the exact column header descriptions in the database schema. In such cases, Llama 3.1 70B often converted only the clearly defined parts of the query into RA, leading to incomplete execution and reduced accuracy. On the other hand, CoT+SC exhibits low recall, as it is highly susceptible to confusion by ambiguous query elements. This causes CoT+SC to frequently generate incorrect code that fails execution, significantly lowering its recall rate.

\vspace{0.25cm}
\noindent\fbox{%
    \parbox{0.98\linewidth}{%
        \textbf{RQ2 findings:}
        GateLens demonstrates high robustness in real-world conditions. It effectively filters out-of-scope inputs, achieving approximately 40\% higher precision than the baseline system. Furthermore, it handles imprecise or informal queries with superior performance, more than doubling the recall (78\% vs. 36\%). This confirms that the system can manage ambiguous user inputs without sacrificing the accuracy of the generated code. 
    }%
}

\subsection{Effectiveness of the RA module (RQ3)}
\begin{figure}[t]
    \centering
    \begin{subfigure}[b]{0.45\linewidth}
        \centering
        \includegraphics[width=\linewidth]{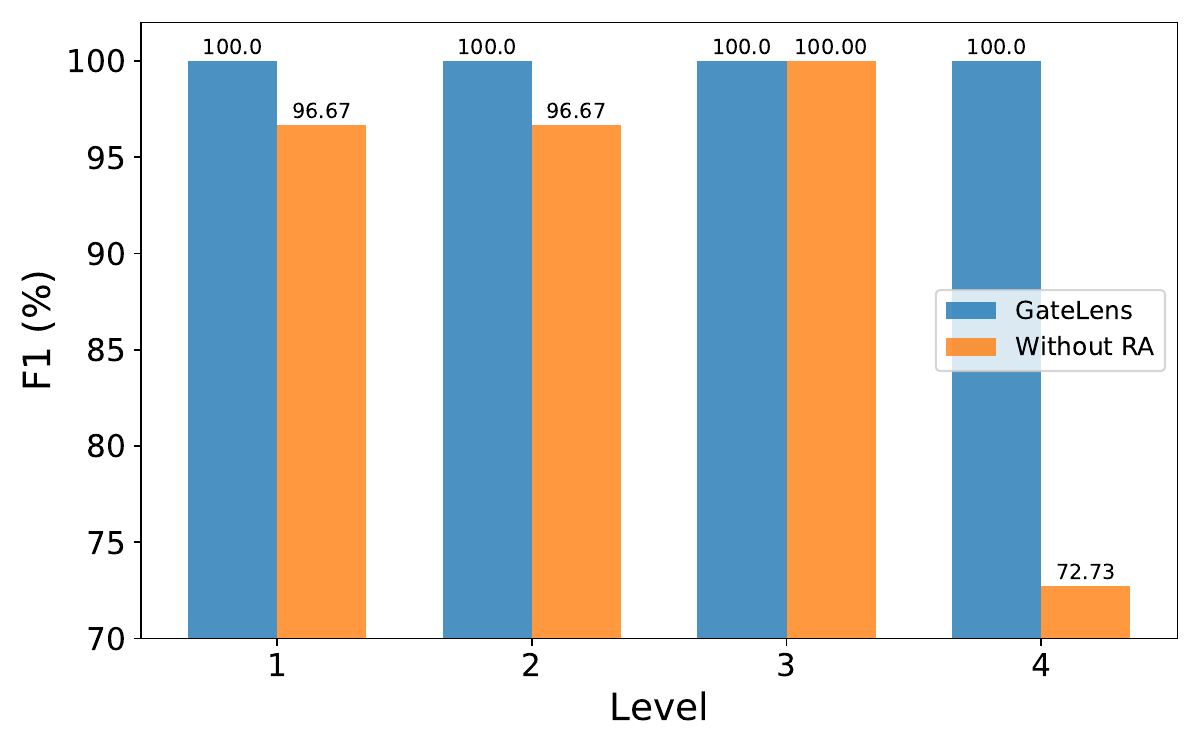}

        \caption{The first benchmark with annotated difficulty levels.}
        \label{fig:compare_without_ra_gt50}
    \end{subfigure}
    \hfill
    \begin{subfigure}[b]{0.5\linewidth}
        \centering
        \includegraphics[width=\linewidth]{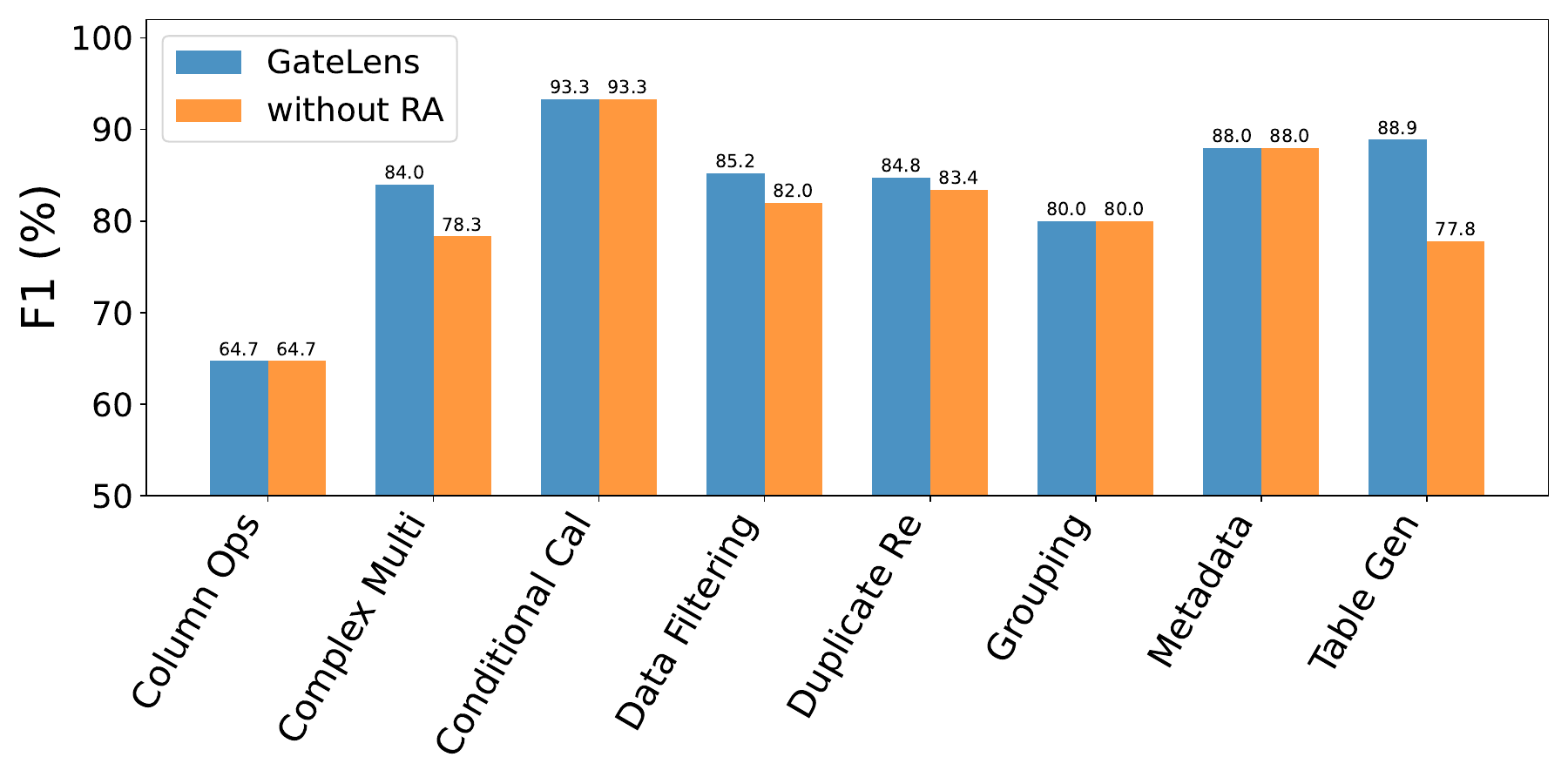}

        \caption{The second benchmark with real-world user queries}
        \label{fig:compare_without_ra_user244}
    \end{subfigure}
    \caption{Comparison of the original method and the method without the RA module across different datasets.}
    \label{fig:compare_without_ra_combined}
 
\end{figure}

To evaluate the impact of the RA module that converts user queries into RA expressions, we conducted experiments by removing the RA module from the framework and comparing the results to the original system. The outcomes, shown in Figure \ref{fig:compare_without_ra_combined}, demonstrate significant performance degradation across both benchmarks when operating without the RA module.

In the first benchmark, performance declined most notably for Level 4 queries, showing a drop exceeding 27\%. These queries, which involve advanced operations like grouping, aggregating, and statistical calculations, demonstrated that RA translation is particularly crucial for handling queries with multiple, intricate operations. Similarly, the second benchmark shows decreased performance in complex tasks such as multi-condition filtering, duplicate data removal, and table generation, further emphasizing RA's effectiveness in managing complex database operations.

The RA module maintained consistent performance for simpler queries, demonstrating its versatility across varying complexity levels. By transforming natural language into precise, logical representations, the RA module serves as a key bridge between user intent and code execution. This translation process enables the code generator to produce accurate, efficient executable code for data analysis tasks.

\vspace{0.25cm}
\noindent\fbox{%
    \parbox{0.98\linewidth}{%
        \textbf{RQ3 findings:}
        The RA module is a critical component for handling query complexity. Removing it results in substantial performance degradation (over 27\% for complex queries), confirming that translating natural language to RA provides necessary structural guidance for accurate code generation.
    }%
}

\subsection{Role of Few-Shot Examples (RQ4)}

\begin{figure}[t]
    \centering  \includegraphics[width=0.6\linewidth]{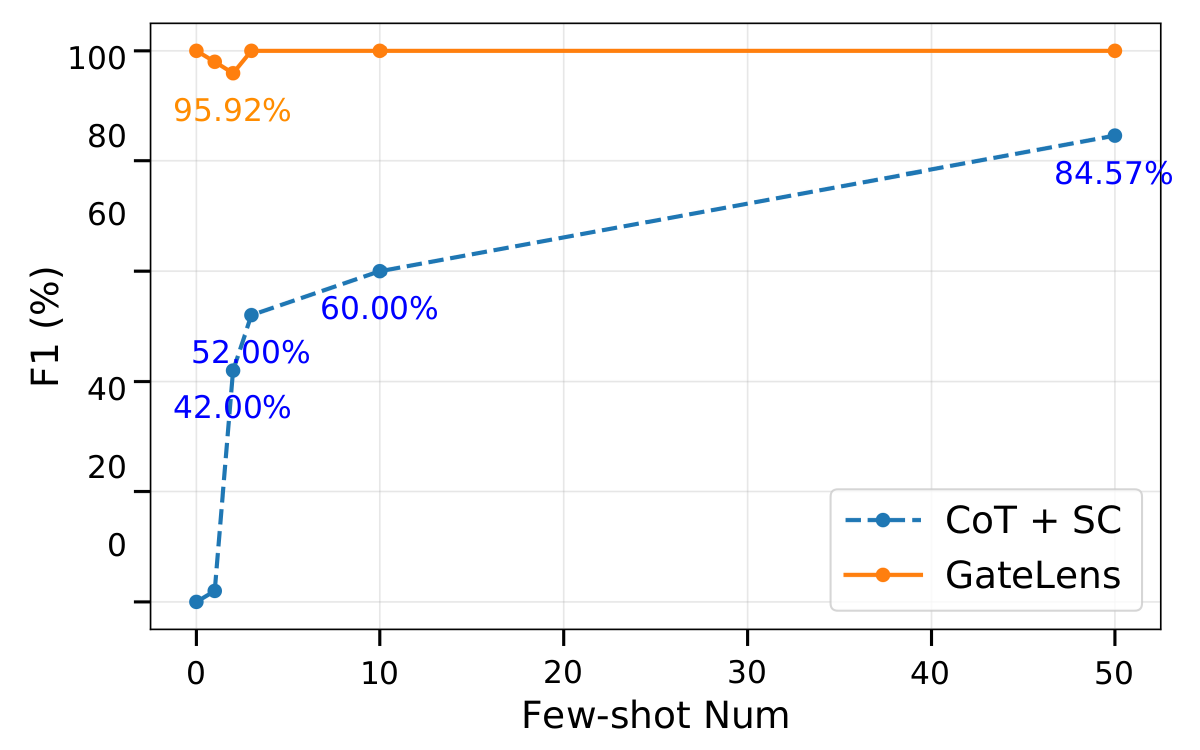}
    \caption{Comparison of GateLens against CoT+SC across different numbers of few-shot examples.}
    \label{fig:few-shot}
\end{figure}

To investigate the effect of including few-shot examples in prompts, we conducted experiments by varying the number of examples provided to both GateLens and CoT+SC. This experiment is performed on the first benchmark containing 50 designed queries, with results illustrated in Figure~\ref{fig:few-shot}.

The results demonstrate that GateLens relies heavily on its RA translation process, achieving 100\% F1 even in a 0-shot setting without any examples. Interestingly, when a small number of examples are added, GateLens becomes slightly biased toward them, leading to a minor degradation in performance (dropping to 95.92\% at 2 examples). However, it regains optimal performance at 3 examples and remains at 100\% thereafter.
In contrast, CoT+SC's performance heavily depends on in-context examples, achieving only 42\% F1 with 2 examples and showing suboptimal results without sufficient few-shot examples. Performance improved steadily with more examples, reaching 84.57\% at 50 examples.

\vspace{0.25cm}
\noindent\fbox{%
    \parbox{0.98\linewidth}{%
        \textbf{RQ4 findings:}
        GateLens does not require many-shot examples to achieve high performance, delivering optimal results in a 0-shot setting. This contrasts with CoT+SC, which depends on carefully curated examples. This independence from few-shot examples makes GateLens more efficient for real-time applications by reducing context size and computational overhead. 
    }%
}

\section{Qualitative Assessment Examples}

\begin{table}[h!]
\centering
\caption{An example of an input query and its corresponding relational algebra expressions translated using a domain-specific database schema, along with the executable code generated from the relational algebra and the resulting output produced when executed over the database.}
\setlength{\tabcolsep}{6pt}
\begin{tabular}{p{3cm}|p{12.4cm}}
\toprule
\textbf{Stage} & \textbf{Content} \\
\midrule
\textbf{Input Query} & \textit{Compute the relative frequency of not run tests for each RC and function} \\
\midrule
\textbf{Relational} & \textbf{Step 1:} $R0 \leftarrow \sigma_{\text{result\_status} = \text{"NotRun"}}(\text{df})$ \\
\textbf{Algebra} & \textbf{Step 2:} $R1 \leftarrow \gamma_{\text{releasecandidate}, \text{function}, \text{countNotRun} \rightarrow \text{count}}(R0)$ \\
& \textbf{Step 3:} $R2 \leftarrow \gamma_{\text{releasecandidate}, \text{function}, \text{totalcount} \rightarrow \text{count}}(\text{df})$ \\
& \textbf{Step 4:} $R3 \leftarrow R1 \bowtie_{\text{R1.releasecandidate} = \text{R2.releasecandidate} \wedge \text{R1.function} = \text{R2.function}} R2$ \\
& \textbf{Step 5:} $\text{Result} \leftarrow \gamma_{\text{R3.releasecandidate}, \text{R3.function}, \text{relativeFrequency} \rightarrow \frac{\text{R3.count}}{\text{R3.totalcount}}}(R3)$ \\
\midrule
\textbf{Python Code} & \begin{lstlisting}[language=Python]
# Step 1: Select rows where result_status is "NotRun"
R0 = df[df['result_status'] == 'NotRun']

# Step 2: Group by releasecandidate and function, count "NotRun" entries
R1 = R0.groupby(['releasecandidate', 'function']).size().reset_index(name='count')

# Step 3: Group all rows by releasecandidate and function, count total entries
R2 = df.groupby(['releasecandidate', 'function']).size().reset_index(name='totalcount')

# Step 4: Join R1 and R2 on releasecandidate and function
R3 = pd.merge(R1, R2, how='inner', left_on=['releasecandidate', 'function'], right_on=['releasecandidate', 'function'])

# Step 5: Compute relative frequency of "NotRun" tests
R3['relativeFrequency'] = R3['count'] / R3['totalcount']
result = R3[['releasecandidate', 'function', 'relativeFrequency']]
\end{lstlisting} \\
\midrule
\textbf{Output} & \begin{tabular}[t]{lll}
\textbf{RC} & \textbf{Function} & \textbf{Freq} \\
ALIEN & func\_001 & 0.833333 \\
ALIEN & func\_002 & 0.142857 \\
ALIEN & func\_003 & 0.125000 \\
ALIEN & func\_004 & 0.142857 \\
PACMAN & func\_001 & 0.325000 \\
$\vdots$ & $\vdots$ & $\vdots$ \\[2pt]
\multicolumn{3}{l}{\small Total rows: 3,662}
\end{tabular} \\
\bottomrule
\end{tabular}
\label{tab:successful}
\end{table}

\begin{table}[h!]
\centering
\caption{An example of an incorrect implementation where the relational algebra translation is semantically correct but the generated code fails to properly implement the projection operation, returning all occurrences of the baseline value rather than the distinct value as specified by the relational algebra.}
\setlength{\tabcolsep}{6pt}
\begin{tabular}{p{3cm}|p{12.4cm}}
\toprule
\textbf{Stage} & \textbf{Content} \\
\midrule
\textbf{Input Query} & \textit{What is the latest baseline on RM-320?} \\
\midrule
\textbf{Relational} & \textbf{Step 1:} $R0 \leftarrow \sigma_{\text{name} = \text{"RM-320"}}(\text{df})$ \\
\textbf{Algebra} & \textbf{Step 2:} $\text{max\_date} \leftarrow \text{MAX}(\pi_{\text{date}}(R0))$ \\
& \textbf{Step 3:} $R1 \leftarrow \sigma_{\text{date} = \text{max\_date}}(R0)$ \\
& \textbf{Step 4:} $\text{Result} \leftarrow \pi_{\text{baseline}}(R1)$ \\
\midrule
\textbf{Python Code} & \begin{lstlisting}[language=Python]
# Step 1: Select rows where the truck name is RM-320
R0 = df[df['name'] == 'RM-320']

# Step 2: Identify the latest date from the selected rows
max_date = R0['date'].max()

# Step 3: Select rows that have this latest date
R1 = R0[R0['date'] == max_date]

# Step 4: Project the baseline field
result = R1['baseline'].reset_index(drop=True)
\end{lstlisting} \\
\midrule
\textbf{Output} & \begin{tabular}[t]{l}
\textbf{Baseline} \\
23\_T2\_w2509\_ABC\_12 \\
23\_T2\_w2509\_ABC\_12 \\
23\_T2\_w2509\_ABC\_12 \\
23\_T2\_w2509\_ABC\_12 \\
23\_T2\_w2509\_ABC\_12 \\
$\vdots$ \\[2pt]
\multicolumn{1}{l}{\small Total rows: 132 (all duplicates)}
\end{tabular} \\
\bottomrule
\end{tabular}
\label{tab:failed}
\end{table}

To provide a closer view into the inner workings of the GateLens framework, we analyze two representative examples. Each example presents the relational algebra solution generated from the input query using a domain-specific relational model, the corresponding executable code created from the relational algebra statements, and the resulting output after execution.

Table~\ref{tab:successful} represents an example that produces the desirable result. The agent correctly decomposes the input query into a sequence of relational algebra statements that include filtering, grouping by release candidate and function, counting entries with the status “NotRun,” joining the intermediate results, and finally computing the relative frequency. This process shows how the system is capable of understanding the semantics of the query, applying the required relational transformations, and leveraging the domain-specific schema to produce the correct result. This stepwise reasoning lays the foundation for semantic decomposition and precise query interpretation, which then naturally translates into valid executable code.

Table~\ref{tab:failed} is an example that illustrates a failed case. Although the relational algebra translation is semantically sound, the resulting code does not correctly apply projection—it returns all occurrences of the same baseline value rather than the distinct value expected by the relational algebra. This issue arises because the plan omits the duplicate-elimination operator ($\delta$), which should be used to ensure the final result contains unique entries. As a result, while the output does not match the intended result, it is still potentially useful to the end user, as it contains relevant baseline information.

Overall, these examples highlight how the reasoning-enhanced LLM agent supports accurate and explainable code generation by decomposing user queries into relational algebra operations that align closely with domain-specific requirements. The relational modeling serves as a crucial guide for transforming natural language queries into semantically coherent, executable plans.

\section{Industrial Deployment: Lessons Learned}
\label{sec:lessons}
\begin{table*}[h!]
\small
\centering
\caption{GateLens (zero-shot) vs CoT+SC (few-shot) performance across different roles on the second benchmark. For each role tested, CoT+SC was trained using examples from the other two roles only (leave-one-role-out approach).}

\resizebox{1\textwidth}{!}{
\begin{tabular}{|l|c|c|c|ccc|}
\hline
\multirow{2}{*}{\textbf{Roles}} & \multirow{1}{*}{\textbf{\# Queries}} & \textbf{GateLens}  & \textbf{CoT+SC (Few-shot with All Roles)} & \multicolumn{3}{c|}{\textbf{CoT+SC (Few-shot with Leave-One-Role-Out)}} \\
\cline{3-7}
                   &  \textbf{(244 in total)} & \textbf{{F1 Score}} & \textbf{{F1 Score}} & \textbf{{Without Mechanic}} & \textbf{{Without Project}} & \textbf{{Without Software}} \\ \hline
Mechanically-oriented         & 36                              & {94.25\%}   &  {\textcolor{blue}{80.60\%}}     & 73.85\% {\textcolor{blue}{$\downarrow$ (-6.75\%)}}           & 86.57\%                & 80.60\%           \\ \hline
Project-oriented              & 193                          & {80.21\%}      &   {\textcolor{blue}{78.93\%}}  & 78.93\%                   & 76.22\% {\textcolor{blue}{$\downarrow$ (-2.71\%)}}        & 78.40\%           \\ \hline
Software-oriented             & 15                            & {100\%}         &   {\textcolor{blue}{100\%}}  & 100\%                   & 100\%                   & 82.76\%  \textcolor{blue}{$\downarrow$ (-17.24\%)}   \\ \hline
% \textbf{Total}                & \textbf{244}     & \textbf{86.02}          & \textbf{81.14}       & \textbf{83.15}   &    -      &    -    &  - \\ \hline
\end{tabular}
}

\label{tab:combined_performance}
\end{table*}

The deployment of GateLens at a partner automotive company has provided valuable insights into integrating AI-assisted analytics into complex industrial workflows, specifically for streamlining decision-making in automotive software release validation.

%The integration process reflects the inherent complexity of the automotive domain, where a vehicle is constructed through three primary stages of integration: subsystem (control unit) level, system (multiple control units) level, and the complete vehicle level. At each of these levels, extensive testing is conducted, and the results are logged into a central database. The Go/No-Go decisions---critical milestones in determining whether a release is approved---are made at every integration level. However, stakeholders from diverse backgrounds—ranging from project management and mechanical engineering to software engineering—need to query this raw database to assess product quality and make informed release decisions. Many of these stakeholders, particularly those with project or mechanical orientations, lack expertise in data analytics, leading to inefficiencies in decision-making.
Automotive software integration at the company typically occurs across three hierarchical stages: subsystem (control unit), system (multiple control units), and full vehicle levels. Each stage involves extensive testing, with results stored in a central database. Critical Go/No-Go decisions are made at these stages to determine whether a release meets quality thresholds. However, stakeholders from diverse backgrounds---including project managers, mechanical engineers, and software engineers---must query the raw data to evaluate product quality. Many lack expertise in data analytics, creating bottlenecks and delays in the decision-making process.

%Previously, the analytics for Go/No-Go decisions relied on a small team of 2-3 Full-Time Equivalents (FTEs), often overwhelmed by the vast number and diversity of queries. To handle the workload appropriately, the team size would need to triple. This is the effort that GateLens effectively automates, allowing decisions to be made more efficiently and accurately. Currently, GateLens is in an extended pilot phase, with access granted to a pool of 60-80 users. The analytics team has shifted into a support role, helping stakeholders translate their broad needs into a set of precise prompts for GateLens, thereby bridging the gap between technical complexity and user requirements.
Previously, these analytics were managed by a small team of 2–3 full-time analysts, who were often overwhelmed by the volume and diversity of requests. Scaling the team to meet the current demand would have required tripling its size. GateLens addresses this challenge by automating much of the workload, enabling more efficient decision-making. Currently, GateLens is in an extended pilot phase, supporting a pool of 60-80 users. The analytics team has transitioned to a support role, helping stakeholders articulate their needs into clear, actionable prompts for the system.
%Several lessons were learned during the deployment process. 
%First of all, user acceptance has been a phased process, beginning with a small-scale pilot by the data analytics team to establish the initial benchmarks. This was followed by a second round of piloting that included five additional users from diverse backgrounds, leading to the creation of an expanded benchmark. The current phase involves a much larger group of 60-80 users, with initial feedback being highly encouraging. 
User adoption of GateLens has progressed in phases:

\begin{itemize}

\item \textbf{Small-Scale Pilot}: The initial deployment within the analytics team established benchmarks.

\item \textbf{Expanded Pilot}: Five additional users from varied backgrounds contributed to refining the benchmarks.

\item \textbf{Wider Rollout}: The current phase involves a larger group of 60–80 users. Feedback has been highly positive, with stakeholders recognizing GateLens’s ability to simplify and accelerate complex analyses.
\end{itemize}

Since the launch, the number of both new and recurring users has grown, encompassing diverse roles and types of queries, thereby demonstrating the tool’s increasing utility and trust. GateLens significantly reduces the time and effort required for complex analyses, but the shift towards automation also requires users to take on more responsibility in defining and clarifying their needs. The transition from a primarily supportive tool to a more fully automated system is ongoing, demanding a gradual approach with careful calibration to ensure the tool continues to meet evolving needs.

As the system was opened to a broader audience, the diversity of query types increased substantially. While the initial CoT+SC-based agent performed well for a relatively homogeneous user group, its performance became increasingly sensitive to the coverage of few-shot examples. Approaches that rely heavily on few-shot prompting are inherently constrained by example selection and may struggle with previously unseen query patterns. This sensitivity limits their scalability in dynamic industrial environments where new query types continuously emerge. For this reason, we prioritized architectural choices that improve robustness and generalization across roles and query styles.

%To focus on this aspect, we categorized the results from our experiments conducted in Section~\ref{subsec:rq1} (on the second benchmark) into three diverse groups based on their roles in the company and corresponding query patterns and analytical needs: mechanically oriented, project oriented, and software oriented. The system's performance varied across these groups, reflecting the diversity in their needs and query patterns. Mechanically oriented roles exhibited the highest precision (100\%) but lower recall (79.31\%), indicating that while responses were accurate, some relevant queries were mishandled.  In contrast, software-oriented roles saw the strongest overall performance. For project-oriented roles (the largest group), the system achieved a balanced performance across precision and recall. These results emphasize the variability in system performance across stakeholder groups and the need to build generalized solutions to meet diverse requirements.
To explore the system's generalizability, we categorized the roles  within the company % from our experiments in Section~\ref{subsec:rq1} (on the second benchmark) 
into three groups: mechanically-oriented, project-oriented, and software-oriented roles. 
% These roles were distinguished by their query patterns—
%The queries from these roles have distinct focuses.
Mechanically-oriented roles typically focus on truck-specific data filtering. Project-oriented roles often combine meta-queries with conditional filters for release management and statistical analysis. Software roles emphasize truck software applications and user functions.
We can see from Table~\ref{tab:combined_performance}, both GateLens (zero-shot) and CoT+SC (few-shot) exhibit differences in system performance across these groups, which is likely stemming from the complexity and variety of their typical queries. Nevertheless, the results demonstrate that GateLens is capable of supporting all groups to a high degree. To further evaluate CoT+SC's dependency on few-shot examples, we conducted a \textbf{leave-one-role-out} experiment. In this approach, examples from a specific role are excluded in each iteration. For instance, 'without software' indicates that all examples from the software-oriented role have been removed, while the total number of examples is maintained by substituting them with examples from other roles. This highlights the potential challenges with the robustness and generalizability of techniques that rely on few-shot examples. This is a crucial factor to consider in industries where diverse teams collaborate and a wide range of queries may arise.

The impact of automated systems, such as GateLens, on the release process has been substantial. 
Compared to the previous manual process, GateLens has reduced the time required for Go/No-Go analytics by more than 80\%, significantly improving operational efficiency. %This figure is derived from a before-and-after comparison of person-hours per release—shifting from five full-time analysts ($\approx$200 hours) to a single consulting analyst ($\approx$40 hours) per cycle. 
Crucially, the reported 80\% reduction reflects an operational end-to-end improvement that includes the time engineers spend verifying system outputs. In safety-critical industrial environments, AI-generated results are never accepted without validation; therefore, laboratory correctness metrics alone are insufficient to assess real-world impact.
A traditional LLM pipeline that directly translates a natural-language query into executable code and returns a final result is not practical in this context. When such black-box outputs are cross-checked against existing dashboards—a natural and common validation strategy—any discrepancies become difficult to diagnose. If the reasoning process is opaque, engineers cannot trace the source of the error, which directly undermines trust and limits usability. On the other hand, when the reasoning process is expressed in natural language, it becomes possible to follow where things went wrong, but precise intervention remains difficult because natural language is inherently fuzzy and imprecise.

In contrast, GateLens generates an explicit relational algebra (RA) plan that decomposes the input query into logical, stepwise building blocks before producing executable code. This intermediate representation mirrors how experienced developers would structure complex analyses. As a result, engineers can inspect whether the decomposed plan is logically sound before or alongside reviewing the final output. This makes discrepancies diagnosable rather than opaque; the RA-based intermediate representation decreases the effort required for validation and actively builds user trust over time. Consequently, stakeholders can now focus on high-level decision-making, freed from the burden of data preparation and analysis.

%A key takeaway is the value of a domain-specific solution like GateLens, which is easier to understand, debug, and adapt to nuanced internal procedures compared to a more general-purpose data analytics assistant like TaskWeaver. A domain-specific, LLM-based agent can better account for the intricacies of internal workflows and procedures, providing a more effective and reliable solution. The implementation of such tools must be carefully tailored to balance automation with user empowerment, ensuring that stakeholders can leverage the system without being overwhelmed by its complexity.
A key advantage of GateLens lies in its domain-specific design. Unlike general-purpose tools like TaskWeaver~\citep{qiao2023taskweaver} or AutoGen~\citep{wu2023autogen}, GateLens is tailored to automotive workflows, making it easier to understand, debug, and adapt to automotive common procedures. This focus on domain relevance ensures that the system aligns more closely with stakeholders’ needs while providing reliable and nuanced support.

Post-deployment monitoring revealed practical failure modes that highlight challenges of LLM adoption in industrial settings. Most issues did not stem from complex reasoning errors, but from ambiguities in user queries. The two most common cases were: (i) implicit constraints, where users assumed a specific test environment or time frame without stating it explicitly, and (ii) highly localized team jargon that led to incorrect schema mappings. To address these issues, we systematically documented recurring patterns, refined prompt guidelines, and extended the schema metadata with an explicit jargon glossary, incorporated as domain-specific context to improve term-to-column alignment. This process also repositioned the central analytics team toward supporting clearer, more explicit query formulation.

%In summary, the deployment of GateLens in the automotive software release validation process highlights both the opportunities and challenges of integrating AI-driven tools into complex workflows. The system has improved decision-making efficiency and operational processes, achieving encouraging user acceptance. However, its success relies on ongoing refinements and careful management of the transition to full automation, especially in an industry as intricate as automotive, where diverse stakeholder needs must be addressed. GateLens demonstrates how domain-specific AI solutions can transform critical processes in the automotive sector and beyond.
In summary, the deployment of GateLens demonstrates how domain-specific AI solutions can transform critical workflows in the automotive sector. By automating labor-intensive processes and enhancing decision-making, GateLens has delivered measurable improvements in efficiency and user satisfaction. However, its success depends on ongoing refinement and careful management of the transition to full(er) automation. 
%RF: I'm not sure we should write as if full automation is the vision and end goal, maybe better to really be a super-advanced "partner" to the senior analysts and team members?
Balancing automation with user empowerment remains crucial, particularly in a complex industry like automotive, where diverse stakeholder needs must be met.

GateLens represents a promising step forward, showcasing the potential of AI-driven systems to improve not only the automotive domain but also other industries requiring robust, scalable solutions for intricate processes.

\section{Related Work}
\label{sec:related_work}
General-purpose LLMs are primarily designed for and trained on natural languages. Working with tabular data requires specialized adaptations to effectively handle its structured and heterogeneous nature~\citep{fang2024large, wang2024chain, van2024tabular}. First, the structured tabular data is typically transformed into serialized text. The performance of the LLM may depend on this transformation~\citep{min2024exploring}. Subsequently, the serialized text data is used as input to the LLM for various tasks, such as question-answering, summarization, or logical reasoning. Common approaches to improve LLM performance include prompt engineering, pre-training, fine-tuning, and Retrieval-Augmented Generation (RAG).

Pre-training and fine-tuning~\citep{zhang2024tablellama, parthasarathy2024ultimate, vm2024fine, dong2022table, hegselmann_tabllm_nodate} often face scalability concerns. Although resource-efficient training techniques have been proposed to mitigate the substantial computational demands of LLMs~\citep{han2024parameter, 10.1145/3626772.3657807}, in safety-critical applications with evolving data and requirements, training LLMs presents significant challenges due to the constant need for rigorous validation and verification. This ongoing necessity substantially increases resource demands for development and maintenance, potentially exceeding the capacities of many companies. Techniques such as RAG have been employed to dynamically integrate external knowledge bases during inference, reducing the need for frequent model updates~\citep{zhao2024retrieval, gao2023retrieval}. However, such methods can pose challenges in safety-critical industrial settings as well, since both retrieval modules and model components must undergo synchronized updates to maintain relevance, reliability, and compliance with validation and verification requirements.
Costs would also be especially high with fine-tuning since re-tuning would be needed when new and improved base LLMs are released and should be incorporated.

Prompt engineering techniques are among the most resource-efficient methods for improving LLM output~\citep{sahoo2024systematic, jin2023tab}. From a user standpoint, when the input is natural language, prompting techniques can be broadly categorized based on the type of language generated by the LLM. These include outputs in natural language, structured languages~\citep{li2023chain}, or symbolic languages. When the generated language is natural language, LLMs often fail to consistently follow instructions, particularly when the instructions are complex or require precise, step-by-step execution~\citep{pham2024suri}. This inconsistency arises because natural language, while flexible and expressive, can be ambiguous and prone to misinterpretation by LLMs. Structured languages include general-purpose languages (e.g. Python)~\citep{ye_dataframe_2024}, query languages (e.g. SQL)~\citep{li_can_nodate, dong2023c3, mouravieff-etal-2024-learning}, configuration formats (e.g. YAML or JSON), or other Domain-Specific Languages (DSLs)~\citep{glenn2024blendsql, dai2024uqe}. These languages are subsequently interpreted and/or executed by either external tools, the same LLM, or another LLM agent. This approach offers significant advantages, as it enables precise execution of tasks. Another popular type of output is symbolic languages. Literature shows that symbolic representations provide a more rigorous framework for articulating premises and intent, which can enhance reasoning capabilities~\citep{pan2023logic}. 

In this paper, we introduce a novel prompt-only (training-free) approach that bridges natural language and executable code through RA, a symbolic formalism designed for relational modeling and ideally suited for analyzing tabular data. Unlike prior work that often relies on complex multi-agent planning, our approach leverages RA as a lightweight intermediate representation to enable precise query normalization, disambiguation of natural language input, and efficient code generation.
RA acts as an abstraction layer that can target multiple execution backends (e.g., Python, SQL), providing adaptability across systems. In GateLens, we generate Python code to support practical industrial deployment and high-performance execution. GateLens is \emph{training-free}, \emph{feed-forward} (single-pass, without looping or multi-agent orchestration), and thus easier to verify, trace, maintain, and trust -- qualities critical for safety-critical industrial applications.

\section{Discussion and Conclusions}
%This study introduced GateLens, a reasoning-enhanced LLM agent specifically designed for automotive software release validation. GateLens demonstrates significant advancements by reducing analysis time by over 80\% while maintaining high accuracy in test result interpretation, impact assessment, and release candidate evaluation. The inclusion of a Relational Algebra reasoning module is a critical factor in improving robustness and scalability, as evidenced by superior F1 scores in both benchmarking and industrial evaluations.

This study introduced GateLens, a reasoning‑enhanced LLM architecture for reliable tabular analysis, applied to domain‑specific software release validation in the automotive industry. % While the core challenge can be related to text-to-SQL generation, GateLens introduces a novel perspective by employing RA as an intermediate representation before code generation. 
By introducing RA as an intermediate representation before code generation, GateLens addresses the ``Unfaithful Chain-of-Thought (CoT) reasoning'' problem in code generation, where reasoning steps in CoT explanations do not accurately reflect the model's actual thought process~\citep{turpin2023language}. Specifically, GateLens divides the analysis into two steps: (1) natural language queries are first translated into RA expressions, and (2) these RA expressions are then converted into executable code. We use Python as our target language in step (2) due to its widespread use in our partner company and the model’s strong performance in Python, which benefits from more extensive training data. However, this step is also compatible with RA-to-SQL generation, enabling flexibility across backends. The inclusion of the RA reasoning module is a critical factor in improving robustness and scalability, as evidenced by superior F1 scores in both benchmarking and industrial evaluations.
% By dividing the process into two explicit steps—intermediate RA conversion followed by executable code generation from RA expressions—GateLens increases the likelihood that the model follows intermediate reasoning steps, resulting in more reliable and interpretable code generation.

% The choice of Python over direct SQL generation is motivated by LLMs’ superior capabilities in Python due to the prevalence of Python training data compared to SQL, while SQL generation often requires intricate prompt design or fine-tuning to handle different database dialects. The RA abstraction layer provides adaptability across execution backends while maintaining the training-free architecture critical for industrial applications.

Our findings regarding few-shot learning (RQ4) highlight a notable architectural advantage of GateLens over conventional few-shot learning approaches. While CoT+SC's performance improves with more examples, this approach introduces several practical and technical challenges. Increasing example count expands input context size, which increases inference time and computational cost due to the quadratic complexity of Transformer-based models—particularly problematic for real-time and resource-constrained applications~\citep{cui2025stepwise, agarwal2024many}. This also escalates operational costs through increased token usage (higher cloud API fees) and computational demands~\citep{cui2025stepwise}. Additionally, the quality and selection of examples significantly impact performance~\citep{huang2023fewer}; poorly chosen or noisy examples can degrade reasoning and lead to overfitting or failure to generalize on complex tasks~\citep{qin2024context, zhou2024can}. As more examples are added, the risk of including irrelevant or contradictory rationales amplifies, potentially confusing the model and reducing accuracy~\citep{cui2024theoretical, zhou2024can}. Moreover, the larger context risks exceeding the maximum length, which can lead to truncation or lost information~\citep{agarwal2024many}, compromising the model's response quality. In long contexts, critical content may be ignored, resulting in a phenomenon known as ``lost in the middle,'' which adversely affects overall performance~\citep{liu2023lost}. Finally, crafting high-quality, task-specific CoT examples is labor-intensive~\citep{tai2023exploring, stechly2024chain}, and while many-shot in-context learning (ICL) may improve performance on specific tasks, generalization to new tasks remains limited without careful prompt engineering. GateLens circumvents these issues by achieving optimal performance in a zero-shot setting, relying on logical RA translation rather than using manually designed many-shot examples. By maintaining a compact input footprint, GateLens reduces average total token consumption by approximately 79\% compared to the CoT+SC baseline in our production logs, leading to proportionally lower inference latency and API usage costs.

While our lab-based quantitative metrics provide a critical, controlled baseline for comparing architectural choices and measuring robustness against ground truths, we acknowledge that precision and recall on curated queries cannot guarantee absolute correctness in open-ended deployment. Therefore, these lab validations serve primarily to complement—rather than replace—our industrial evaluation.
In real-world deployment serving 60-80 users, GateLens demonstrates significant practical value through its user-friendly interface and robust query processing capabilities, with users particularly appreciating the flexibility to input, debug, and refine queries easily. This marks a substantial advancement in industrial data interaction, successfully handling complex and ambiguous queries while providing practical support for faster decision-making in safety-critical software release processes. GateLens demonstrates significant practical advancements by reducing analysis time by over 80\% while maintaining high accuracy in test result interpretation, impact assessment, and release candidate evaluation. 

Our implementation insights highlight the advantages of focusing on training-free and single-pass agent systems by foregrounding the perception phase in the code generation pipeline. This modular architecture opens opportunities for incorporating emerging LLM capabilities while preserving the system's practical utility in safety-critical industrial applications. A phased deployment program is ongoing and shows that multiple stakeholder groups can be supported, though the evolving roles and analytical needs require continued refinement.

Although instantiated in the automotive release domain, the GateLens architecture can be applied to other domains with comparable analytical requirements. Its core components---the RA-based intermediate reasoning layer, the separation between query interpretation and code generation, in-scope validation, and zero-shot operation without reliance on few-shot examples---do not depend on automotive-specific properties. Adapting the system to a new domain primarily requires replacing the schema and domain knowledge base, while preserving the architectural reasoning pipeline. Domains characterized by structured tabular data, complex stakeholder queries, and safety- or compliance-critical decision-making (e.g., healthcare analytics, financial auditing, or certification workflows) share these properties, suggesting broader applicability.

Future work will focus on validating this architectural transferability and testing alternative LLM configurations to enhance reliability across other safety-critical industries. By bridging the gap between flexible natural language interaction and rigorous analytical standards, this approach demonstrates the potential for reasoning-enhanced LLMs to transform industrial workflows across a broad spectrum of critical applications.

%Future work will address current limitations by expanding domain applicability beyond automotive software and testing alternative LLM configurations to enhance reliability across other safety-critical industries. This approach demonstrates the potential for LLM-based solutions to transform industrial data analysis workflows while maintaining the reliability standards required for critical applications.

\section{Threats to Validity} 

The validity of our findings is subject to several potential threats. First, this framework depends on well-defined, static schemas for accurate query decomposition, which fundamentally limits its effectiveness in environments with incomplete or frequently changing schemas. Second, the benchmarks and query scenarios used for evaluation, derived from historical data and real-world queries, may not fully capture the diversity and complexity of potential use cases, which could impact the robustness of the system in broader deployments. Finally, the system’s performance depends on the specific LLM configurations used, such as GPT-4o and Llama 3.1 70B, and their ability to interpret and generate RA expressions. Future work will address these limitations through broader domain testing, expanded evaluation scenarios, and alternative LLM configurations.

\section{Acknowledgement}
This work was partially supported by the Wallenberg AI, Autonomous Systems and Software Program (WASP), funded by the Knut and Alice Wallenberg Foundation, and by the Gender Initiative for Excellence (Genie) at Chalmers University of Technology, funded by the Chalmers University Foundation.

%% If you have bib database file and want bibtex to generate the
%% bibitems, please use
%%
\bibliographystyle{elsarticle-num} 
\bibliography{refs}

%% else use the following coding to input the bibitems directly in the
%% TeX file.

%% Refer following link for more details about bibliography and citations.
%% https://en.wikibooks.org/wiki/LaTeX/Bibliography_Management

%\begin{thebibliography}{00}

%% For numbered reference style
%% \bibitem{label}
%% Text of bibliographic item

%\bibitem{lamport94}
%  Leslie Lamport,
%  \textit{\LaTeX: a document preparation system},
%  Addison Wesley, Massachusetts,
%  2nd edition,
%  1994.

\end{document}